Letter

# Plasmonic Microbubble Dynamics in Binary Liquids

Xiaolai Li, Yuliang Wang,* Binglin Zeng, Marvin Detert, Andrea Prosperetti, Harold J. W. Zandvliet,* and Detlef Lohse*

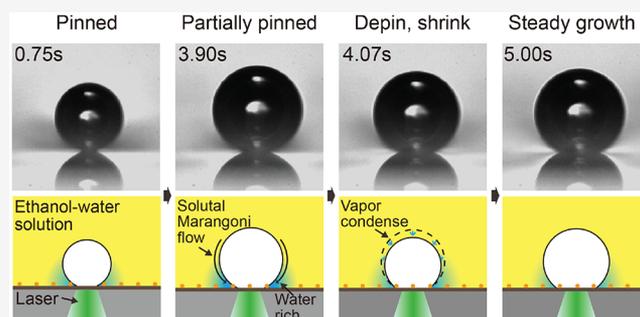

**ABSTRACT:** The growth of surface plasmonic microbubbles in binary water/ethanol solutions is experimentally studied. The microbubbles are generated by illuminating a gold nanoparticle array with a continuous wave laser. Plasmonic bubbles exhibit ethanol concentration-dependent behaviors. For low ethanol concentrations ($f_e$) of ≲67.5%, bubbles do not exist at the solid−liquid interface. For high $f_e$ values of ≳80%, the bubbles behave as in pure ethanol. Only in an intermediate window of 67.5% ≲ $f_e$ ≲ 80% do we find sessile plasmonic bubbles with a highly nontrivial temporal evolution, in which as a function of time three phases can be discerned. (1) In the first phase, the microbubbles grow, while wiggling. (2) As soon as the wiggling stops, the microbubbles enter the second phase in which they suddenly shrink, followed by (3) a steady reentrant growth phase. Our experiments reveal that the sudden shrinkage of the microbubbles in the second regime is caused by a depinning event of the three-phase contact line. We systematically vary the ethanol concentration, laser power, and laser spot size to unravel water recondensation as the underlying mechanism of the sudden bubble shrinkage in phase 2.

Plasmonic microbubbles formed around metal nano­particles immersed in liquid and irradiated by a resonant light source have numerous applications ranging from biomedical diagnosis and cancer therapy[1−4] and micro­manipulation of micro/nano-objects[5−7] to locally enhanced chemical reactions.[8,9] A proper understanding of the plasmonic microbubble dynamics is crucial for the optimal usage of these bubbles in the aforementioned applications.

In a previous study, we have shown that plasmonic microbubbles in pure (monocomponent) liquids undergo four different phases: the initial giant vapor bubble (phase I), the oscillating bubble (phase II), water vaporization-dominated growth (phase III), and air diffusion-dominated growth (phase IV).[10] The bubble dynamics in these different phases depends on several factors, such as particle arrange­ment,[7] laser power, gas concentration,[10−12] and types of liquid.[13,14] The nucleation of a giant bubble is initiated by locally heating the liquid around the plasmonic particle up to the spinodal temperature.[10] The heating process is limited by thermal diffusion and can theoretically be described well.[15−17] Moreover, the dissolved gas plays a major role in bubble nucleation: an increase in the relative gas concentration leads to a lower bubble nucleation temperature.[10,11] The radius of the bubble in air diffusion-dominated phase IV is $R(t) \propto t^{1/3}$ for air equilibrated or air-oversaturated water.[11,12] Further­more, several studies of the shrinkage of bubbles have revealed that the shrinkage depends on the illumination history of the bubble as well as the gas concentration level of the liquid.[18,19]

Most of these studies were conducted in water or other pure liquids,[13] such as *n*-alkanes.[14] However, it is known that many plasmonically assisted processes also take place in binary liquids.[8] The dynamics of plasmonic bubbles in binary liquids has not yet been investigated in detail. Only in a recent study did we start to investigate the formation of the initial giant plasmonic bubbles in binary liquids.[20] It was shown that the dynamics of bubble nucleation depends on the exact composition of the binary liquid, which could be rationalized on the basis of the constituents of the binary liquid exhibiting different physicochemical properties, such as boiling point and latent heat. In the aforementioned study, the focus was only on phase I of the plasmonic bubble formation process, namely, bubble nucleation. A thorough and complete understanding of the full long-term growth dynamics of plasmonic bubbles in binary liquids is still lacking.

In this Letter, we study the growth dynamics of plasmonic bubbles in ethanol/water mixtures over a large range of parameters and over the full bubble lifetime cycle. The bubble dynamics in various ethanol/water mixtures was monitored by two high-speed cameras: one provides a side view and the other the bottom view, both at a frame rate of 1000 fps (see Experimental Methods in the Supporting Information for details).

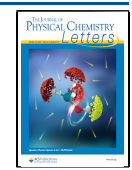

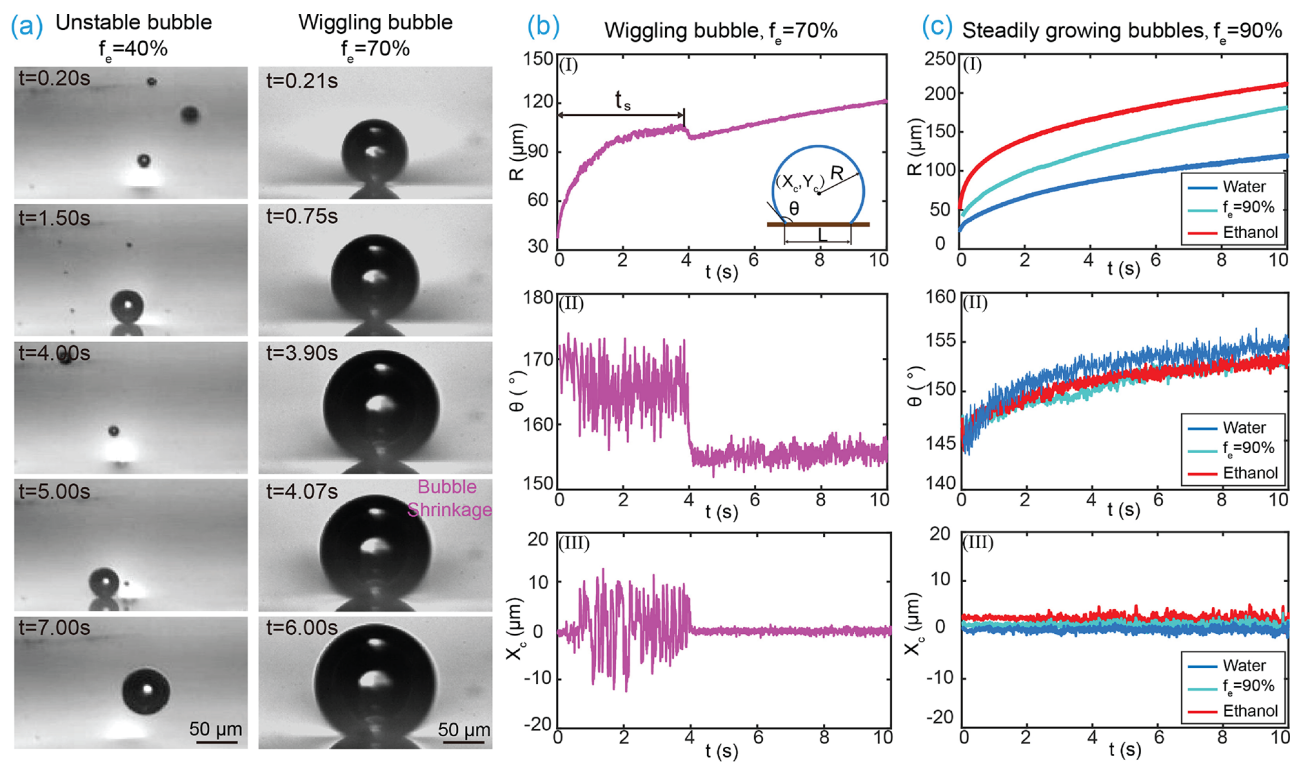

**Figure 1.** (a) Dynamics of plasmonic bubble growth in ethanol/water mixtures with ethanol concentrations of 40% (left) and 70% (right). In the binary liquid in which $f_e$ = 40%, the bubble detaches from the substrate after nucleation, whereas in the binary liquid in which $f_e$ = 70%, the bubble remains pinned on the substrate and gradually grows. A sudden shrinkage is observed at $t$ = 4.07 s. (b) Radius $R(t)$ (I), contact angle $\theta(t)$ (II), and horizontal center position $X_c(t)$ (III) of the bubble in the binary liquid for which $f_e$ = 70%. The inset in panel I shows a sketch, which defines the radius $R$, the contact angle $\theta$, the horizontal center position $X_c$, and the footprint diameter $L$ of the bubbles. (c) $R(t)$, $\theta(t)$, and $X_c(t)$ for a bubble in a binary mixture in which $f_e$ = 90% compared to the cases for pure water and pure ethanol. The laser power ($P_l$) is 100 mW in all cases in panels a–c.

We first report the plasmonic bubble behavior as a function of ethanol concentration $f_e$ (see Figure 1). First, in binary liquids where the ethanol concentrations 0% < $f_e$ ≲ 67.5% (weight), the bubbles are unstable and not pinned at the substrate. As an example, in Figure 1a (left column), we show the bubble dynamics for an ethanol concentration of 40%. In this regime, after nucleation, the bubble detaches from the substrate. Subsequently, a new bubble is generated and then also detaches.

Next, when 67.5% ≲ $f_e$ ≲ 80%, we find that the plasmonic bubble does not detach from the substrate. As an example for this regime, the dynamics for $f_e$ = 70% is shown in Figure 1a (right column). Instead, after an initial growth phase, the plasmonic bubble starts to wiggle in this regime, with a gradually increasing radius $R$ [Figure 1b (I)], a considerably varying contact angle $\theta$ [Figure 1b (II)], and a horizontal center position $X_c$ [Figure 1b (III)]. During the first few seconds after nucleation, the radius, contact angle, and horizontal center position exhibit erratic behavior. At $t$ = 4.07 s after bubble nucleation at $t$ = 0 s, the bubble abruptly stops wiggling and simultaneously starts to shrink for a short time. Afterward, the bubble continues to steadily grow without exhibiting any wiggling. The abrupt stop of wiggling and the simultaneous start of sudden shrinkage always occur in the regime in which 67.5% ≲ $f_e$ ≲ 80%. We define the moment when this occurs as time of shrinkage $t_s$.

Finally, when the ethanol concentration $f_e$ of the binary liquids is >80%, we observed stable bubble growth, like in pure water and pure ethanol. Figure 1c shows the bubble dynamics in pure water, in pure ethanol, and in a binary liquid in which $f_e$ = 90%. The laser power ($P_l$) was maintained at 100 mW for these three measurements. In these three cases, the bubbles grow monotonously, with a gradually increasing $R$ [Figure 1c (I)]. No visible intermediate shrinkage or wiggling [Figure 1c (II and III)] was observed. Moreover, the radii of the bubbles increase with an increasing ethanol concentration. As known from ref 12, for air-saturated water, the observed long-term growth is dominated by gas diffusion. However, for the binary liquids in which $f_e$ ≳ 80%, bubble growth is dominated by vaporization. Because the gas solubility in ethanol increases with temperature, the gas expelled from the liquid evaporation can be dissolved into liquids again, which has been shown in a recent study in n-alkane.[14]

The results presented above demonstrate that the ethanol concentration in water/ethanol mixtures plays a pivotal role in bubble growth. Our interpretation of the observed phenomenon is as follows. In binary liquids with low ethanol ratios ($f_e$ ≲ 67.5%), the detachment of bubbles from the substrate is caused by solutal Marangoni flow.[21] Because the boiling temperature of ethanol is lower than that of water, at the hot substrate it evaporates much faster than water, leading to an ethanol-depleted region near the three-phase contact line (triple line) of the bubbles. The liquid at the top of the bubble is rich in ethanol and thus has a surface tension that is lower than that of the ethanol-depleted region at the bottom of the bubble, resulting in a downward Marangoni flow. This solutal Maragoni flow pushes the bubble upward, which is similar to what has been observed for a jumping droplet in an ethanol/water concentration gradient.[22] As a result, the bubble detaches when the Marangoni flow is strong enough to overcome the pinning of the bubble at the surface.

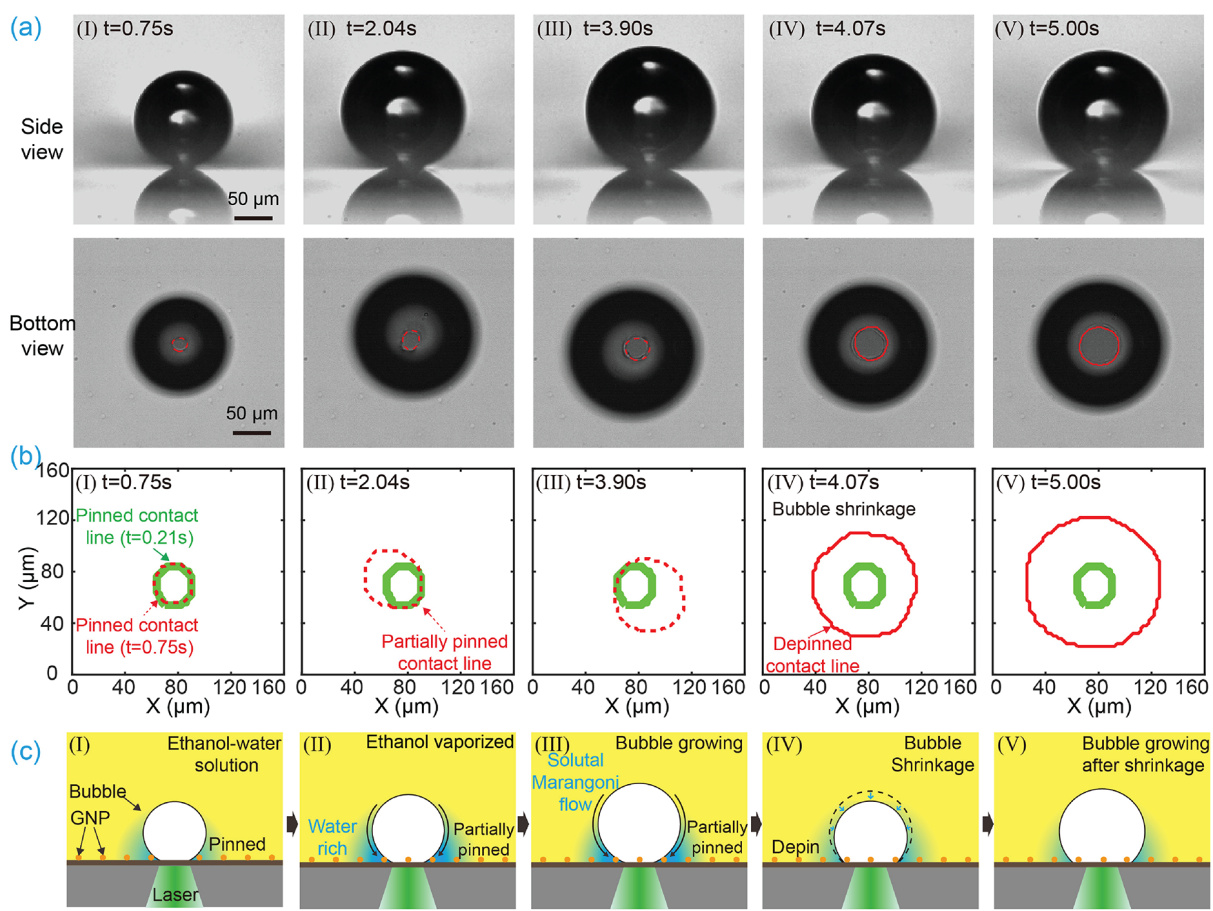

**Figure 2.** Dynamics of bubble shrinkage. (a) Sequential side and bottom view images of a stable bubble in binary liquids in which $f_e$ = 70%. Bubble shrinkage occurs in panel IV. (b) Three-phase contact line of bubbles extracted from the bottom view images of panel a. The three-phase contact line is initially pinned, and later partially pinned, at the edge of the laser spot (green line). At the moment $t_s$ of sudden shrinkage at stage IV, the bubble completely depins. After that, the contact area remains circular with an increasing radius. (c) Schematics of bubble growth and shrinkage in a binary liquid.

The proposed mechanism for bubble detachment can be quantitatively validated. Here we consider an unstable bubble in binary liquids with an ethanol concentration of ≈40−50% as an example. The surface tension difference $\Delta\sigma$ between the bubble top ($f_e \approx$ 40−50%) and bubble bottom (providing $f_e$ = 0−20%) is ≈20−40 mN/m. For a bubble with a radius $R$ of ≈50 μm, the magnitude of the upward Marangoni force $F_M \approx \Delta\sigma R = 1 \times 10^{-6} - 2 \times 10^{-6}$ N. The Marangoni force competes with vertical capillary force $F_p$. Because the measured contact angle of water $\alpha$ = 29.1° (Figure S6a) on our substrate is in the same ballpark as that on a smooth glass surface ($\alpha$ = 24°),[23] the influence of the gold nanoparticles on the pinning/capillary force of the substrate is very limited. As a result, it is reasonable to estimate the vertical capillary force by the equation $F_p \approx \sigma \pi L \sin \theta \approx 1.4 \times 10^{-6}$ N, assuming that the substrate is homogeneous.[24,25] In this expression, surface tension $\sigma$ = 50 mN/m, the medium value of measured contact angle (gas side) $\theta$ = 160°, and the perimeter of the three-phase contact line $\pi L$ = 85 μm.

Notably, the surface tension of the ethanol/water mixtures changes rapidly for small ethanol fractions but remains almost constant for large ethanol fractions[26,27] (see Figure S3). This feature explains the observed peculiar behavior. For pure water, the bubble remains pinned because there is no downward solutal Marangoni flow, but an upward thermal Marangoni flow. For low ethanol concentrations of ≲67.5% due to the strong dependence of the surface tension on $f_e$ in this regime, a large gradient in surface tension can be created and the resulting Marangoni flow is strong enough to depin the bubbles. With an increasing ethanol fraction, the surface tension gradient across the bubble and thereby the Marangoni flow become weaker until finally, at $f_e \approx$ 67.5%, it is unable to overcome the pinning. Therefore, for that ethanol concentration, a transition from detaching bubbles to pinned bubbles with stable growth takes place.

However, this mechanism still does not explain the observed wiggling and sudden shrinkage of the stable bubbles when 67.5% ≲ $f_e$ ≲ 80%. To determine the physical mechanism for the sudden bubble shrinkage, from the bottom view we tracked the changes of the triple line of a bubble during the growth. Figure 2a shows the sequential side view images (first row) and bottom view images (second row) of a bubble in a binary liquid for which $f_e$ = 70%. The bubble shrinks in Figure 2a (IV). In the bottom view images, a dry region (light gray center area) can be observed within the triple line. The dry area becomes circular and increases after the sudden shrinkage (IV → V). Figure 2b shows the profile of the triple line as extracted from the bottom view as well as the spot profile of the laser.

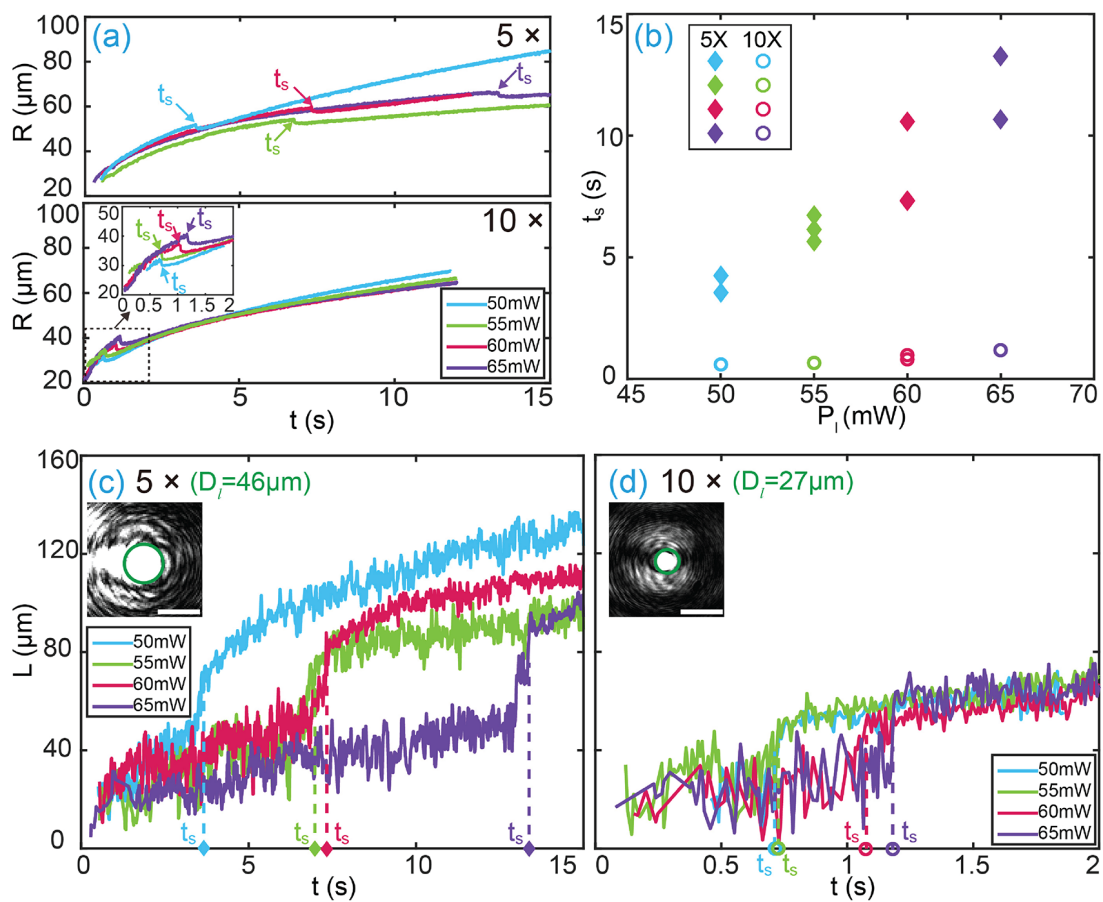

**Figure 3.** Dependence of the bubble shrinkage dynamics on laser power $P_l$ and laser spot size $D_l$ at $f_e$ = 70%. (a) Bubble dynamics for different laser powers $P_l$ at 5× and 10× objectives for laser focusing. (b) Delay time $t_s$ as a function of laser power. (c and d) Footprint diameters of the bubble vs time for 5× and 10× objectives, respectively. The insets show the laser spots. Scale bars of 50 $\mu$m.

It is found that the triple line is first pinned at the edge of the laser spot [first column of Figure 2a (I)] and then partially depins [Figure 2a (II and III)]. This explains the wiggling behavior of the bubble shown in Figure 1b. When the contact line depins completely from the laser spot region [Figure 2b (IV)], the bubble shrinks. Afterward, the triple line remains circular and gradually expands, as shown in Figure 2b (V).

The whole process of bubble growth and shrinkage is schematically shown in Figure 2c. In the early stage of bubble growth, the evaporation of the more volatile ethanol in the binary liquids governs bubble growth. As a result, the water concentration around the three-phase contact line, of which the size is defined by the laser spot, is relatively higher than that of the bulk binary liquid. At the triple line, the rapidly increasing bubble volume tends to expand the bubble and thus moves the triple line outward. At the same time, the higher water concentration around the triple line (due to ethanol evaporation) leads to a locally increased surface tension, resulting in an increased receding contact angle.[28,29] This helps to make the triple line stay pinned. The two effects compete. When the increased surface tension dominates, the triple line remains pinned. Because of the continuous evaporation of the surrounding liquids, the bubble keeps growing. Finally, the contact angle (liquid side) reaches its receding value and the contact line begins to depin. During this process, the contact line is first partially depinned. Finally, once the water around the laser spot region is completely evaporated, complete depinning takes place, exhibiting a rapid decrease in the contact angle (gas side).

After depinning of the three-phase contact line, the region of the laser spot on the sample substrate is isolated by the bubble from the liquid. Because of this effect, the light−vapor conversion efficiency significantly decreases and part of the vapor inside the bubble condenses, resulting in a sudden shrinkage of the bubble [Figure 1b (I)]. A numerical calculation (Figure S2) shows that the sample substrate remains non-isothermal in the vicinity of the triple line when rapid shrinkage takes place. This means that once the bubble depins, the temperature will decrease at the triple line, resulting in vapor condensation and hence rapid bubble shrinkage. The phenomenon resembles similar jumps in the contact angle occurring for evaporation[30−34] or dissolution of binary droplets.[35] After shrinkage, the bubble enters a steady growth phase again, but now with a lower optothermal conversion efficiency, during which there is no visible shrinkage happening again.

Note that the bubble shrinkage due to the pinning effect appears only in the binary liquids with intermediate ethanol ratios of 67.5% ≲ $f_e$ ≲ 80%. In general, considering that the pinning force depends on the wettability and the roughness of the surface, the three-phase contact line motion must be expected to move in the "stick−slip" mode,[36−40] consisting of one or more "stick" phases, in which the contact line is pinned. However, in contrast to this expectation, the bubble growth and the contact angle in pure ethanol and pure water seem to exhibit monotonous growth in our experiments, as shown in Figure 1c. We interpret this finding that for the bubbles in the pure liquids, the pinning effect is only minor and not visible, and the "stick" phase duration is so short that with our temporal resolution of 1000 fps it is invisible in the measured dynamical process.

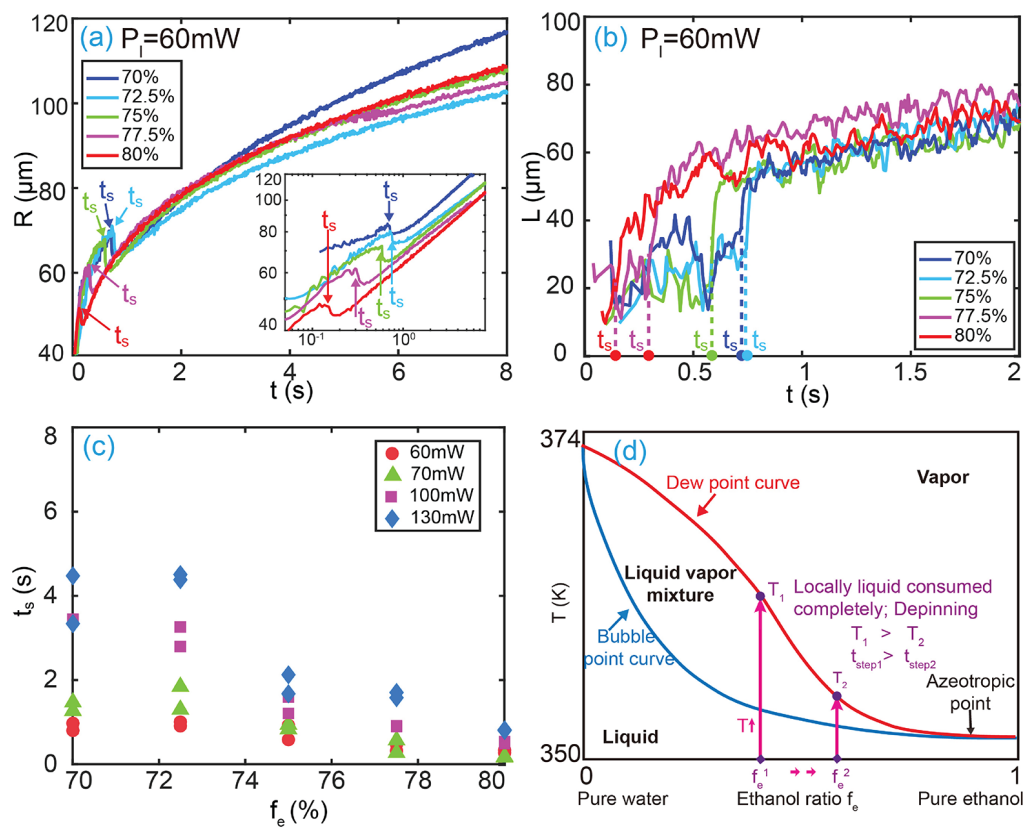

**Figure 4.** Bubble shrinkage dynamics for various ethanol concentrations in the stable bubble regime where 67.5% ≲ $f_e$ ≲ 80%. (a) Radius $R(t)$ and (b) footprint width $L(t)$ at a laser power of 60 mW. (c) Delay time $t_s$ of bubble shrinkage vs ethanol ratio $f_e$ for laser powers of 60, 70, 100, and 130 mW. (d) Schematic diagram of the bubble point and dew point of ethanol/water mixtures, at ambient pressure.

We now further verify the mechanism of bubble shrinkage by varying laser power $P_l$ and the laser spot size. Figure 3a shows the bubble radius versus time at different laser powers for 5× and 10× objectives for laser focusing implying two different laser spot sizes [see panels a and b of Figure S4 for reproducible results of $R(t)$]. In Figure 3b, delay time $t_s$ is plotted versus laser power. For a fixed laser power, the delay time of the 5× objective is substantially longer than that of the 10× objective due to the weaker focusing of the laser beam, resulting in a larger laser spot and a lower power density.

The differences in the observed start moment of shrinkage $t_s$ can be understood by capturing the dynamics of the three-phase contact lines for the 5× and 10× objectives (see panels c and d of Figure 3) (see also Figure S5 for contact line $L$ as a function of radius $R$). After a slight monotonous increase in the footprint diameter of the bubble, it exhibits a sudden jump and eventually gradually increases again. The sudden jump in the footprint diameter coincides with the depinning and shrinkage of the bubble. For the 5× and 10× objectives, the jump in the footprint diameter occurs at ~50 and ~30 μm, respectively. These values agree well with laser spot diameters $D_l$ of 46 and 27 μm for the 5× and 10× objectives, respectively, as shown in the insets of panels c and d of Figure 3. This also confirms that the triple lines are initially pinned by the laser spot and shrinkage occurs because of complete depinning. The laser spot size increases with an increasing laser power. As a result, the heated region and therefore the pinning radius increase. Thus, the delay time $t_s$ is longer for higher laser powers as more liquid needs to evaporate to exceed the pinned region.

Finally, we have investigated the dependence of the bubble shrinkage dynamics on ethanol concentration. As shown in Figure 4a, bubble shrinkage is observed for ethanol concentrations varying from 70% to 80% (see panels c and d of Figure S4 for the reproducible results with different $f_e$ values at laser powers of 60 and 100 mW, demonstrating that this behavior is robust and reproducible). Figure 4b shows the footprint diameter versus time. As we have kept the laser power and laser spot size constant, the footprint diameter at which the sudden shrinkage occurs is almost the same for all ethanol concentrations. Interestingly, the $t_s$ of shrinkage steadily decreases with increasing ethanol concentrations (Figure 4c).

The decrease in $t_s$ with increasing ethanol concentrations is attributed to the fact that the dew point of the ethanol/water mixture decreases with increasing ethanol concentrations (see Figure 4d).[41,42] From the phase diagram, it is immediately clear that at a higher ethanol concentration the ethanol can be evaporated at lower temperatures. Because the laser power and laser spot size are kept constant in this series of experiments, we can safely assume that the heating efficiency remains constant. Therefore, less energy and thus less time are required to evaporate ethanol for a binary liquid with a higher ethanol concentration as compared to a binary liquid with a lower ethanol concentration. The phase diagram also shows that the water/ethanol mixture becomes azeotropic at an ethanol concentration of >90%, implying that the binary liquid starts to behave as a pure liquid. This perfectly agrees with our experimental findings shown in Figure 1c, as beyond $f_e \approx 90\%$ the plasmonic bubble behaves as it does in pure ethanol.

In summary, we have shown that the growth dynamics of plasmonic microbubbles in water/ethanol binary liquids strongly depends on the ethanol concentration. At ethanol concentrations of <67.5%, bubbles nucleate and subsequently detach from the substrate due to the downward Marangoni flow. For ethanol concentrations in the range of 67.5−80%, the bubbles first wiggle. A sudden bubble shrinkage takes place after a delay time $t_s$, originating from the complete depinning of the wiggling bubble from the laser spot area on the sample surfaces. Prior to that event, the bubble was pinned or partially pinned on the laser spot. When the liquid around the pinned region is completely evaporated, instantaneous expansion of the three-phase contact line as well as sudden shrinkage takes place. Finally, when $f_e \gtrsim 80\%$, the binary solutions become azeotropic, leading to a behavior similar to that in pure liquids, and bubbles steadily grow. Knowledge of the sensitive dependence of the plasmonic bubble dynamics on the liquid compositions obtained in this paper may help in exploiting the relevant applications of these bubbles in multiple-component liquids. For instance, when a good local mixing of the binary liquids is preferable, operating in the first regime $f_e \lesssim 67.5\%$, in which the bubble is wiggling, may be preferable. Having understood features is also a prerequisite for exploiting them in optimizing the energy conversion efficiency of plasmonic bubble generation.

## ■ ASSOCIATED CONTENT

### ⓢ Supporting Information

The Supporting Information is available free of charge at https://pubs.acs.org/doi/10.1021/acs.jpclett.0c02492.

> Details of experimental methods, numerically estimated temperature field of the substrate under laser irradiation, surface tension of ethanol/water mixtures versus ethanol concentration in binary mixtures, repeated experimental results of the bubble shrinkage dynamics, dynamics of the three-phase contact line versus bubble radius, and a comparison of bubble shrinkage dynamics on substrates with a smooth gold layer and a gold nanoparticle array (PDF)

## ■ AUTHOR INFORMATION

### Corresponding Authors

**Yuliang Wang** − Robotics Institute, School of Mechanical Engineering and Automation and Beijing Advanced Innovation Center for Biomedical Engineering, Beihang University, Beijing 100191, P. R. China; orcid.org/0000-0001-6130-4321; Email: wangyuliang@buaa.edu.cn

**Harold J. W. Zandvliet** − Physics of Interfaces and Nanomaterials, MESA+ Institute, University of Twente, 7500 AE Enschede, The Netherlands; orcid.org/0000-0001-6809-139X; Email: h.j.w.zandvliet@utwente.nl

**Detlef Lohse** − Physics of Fluids, Max Planck Center Twente for Complex Fluid Dynamics and J. M. Burgers Centre for Fluid Mechanics, MESA+ Institute, University of Twente, 7500 AE Enschede, The Netherlands; Max Planck Institute for Dynamics and Self-Organization, 37077 Göttingen, Germany; orcid.org/0000-0003-4138-2255; Email: d.lohse@utwente.nl

### Authors

**Xiaolai Li** − Physics of Fluids, Max Planck Center Twente for Complex Fluid Dynamics and J. M. Burgers Centre for Fluid Mechanics, MESA+ Institute, University of Twente, 7500 AE Enschede, The Netherlands; Robotics Institute, School of Mechanical Engineering and Automation, Beihang University, Beijing 100191, P. R. China; orcid.org/0000-0002-4912-2522

**Binglin Zeng** − Physics of Fluids, Max Planck Center Twente for Complex Fluid Dynamics and J. M. Burgers Centre for Fluid Mechanics, MESA+ Institute, University of Twente, 7500 AE Enschede, The Netherlands; Robotics Institute, School of Mechanical Engineering and Automation, Beihang University, Beijing 100191, P. R. China; orcid.org/0000-0001-8729-1944

**Marvin Detert** − Physics of Fluids, Max Planck Center Twente for Complex Fluid Dynamics and J. M. Burgers Centre for Fluid Mechanics, MESA+ Institute and Physics of Interfaces and Nanomaterials, MESA+ Institute, University of Twente, 7500 AE Enschede, The Netherlands

**Andrea Prosperetti** − Physics of Fluids, Max Planck Center Twente for Complex Fluid Dynamics and J. M. Burgers Centre for Fluid Mechanics, MESA+ Institute, University of Twente, 7500 AE Enschede, The Netherlands

Complete contact information is available at:
https://pubs.acs.org/10.1021/acs.jpclett.0c02492

### Notes

The authors declare no competing financial interest.

## ■ ACKNOWLEDGMENTS

The authors thank the Dutch Organization for Research (NWO), The Netherlands Center for Multiscale Catalytic Energy Conversion (MCEC), ERC (via Advanced Grant DDD, Project 740479), and the Chinese Scholarship Council (CSC) for the financial support. Y.W. acknowledges the financial support from NSFC (Grants 51775028 and 52075029) and the Beijing Natural Science Foundation (Grant 3182022). The authors also thank Prof. Xuehua Zhang for the fruitful discussions.

## ■ REFERENCES